\documentclass[aps,preprintnumbers,amsmath,amssymb,endfloats]{revtex4} 
\usepackage{dcolumn}
\usepackage{bm}

\pdfoutput=1

   \ifx\pdfoutput\undefined \usepackage[dvips]{graphicx} \else
   \usepackage[pdftex]{graphicx} \pdfcompresslevel=9 \fi
   \usepackage{epstopdf}
   \DeclareGraphicsExtensions{.pdf,.gif,.jpg,.eps,.png}

\begin{document}

\preprint{{\it Preprint}}

\title{The structure of the magnetic reconnection exhaust boundary}
 
\author{Yi-Hsin~Liu}
\affiliation{Los Alamos National Laboratory, Los Alamos, NM 87545}
\author{J.~F.~Drake}
\affiliation{University of Maryland, College Park, MD 20742}	
\author{M.~Swisdak}
\affiliation{University of Maryland, College Park, MD 20742}
\date{\today}

\begin{abstract}
The structure of shocks that form at the exhaust boundaries during
collisionless reconnection of anti-parallel fields is studied using
particle-in-cell (PIC) simulations and modeling based on the
anisotropic magnetohydrodynamic equations. Large-scale PIC simulations
of reconnection and companion Riemann simulations of shock development
demonstrate that the pressure anisotropy produced by counterstreaming
ions within the exhaust prevents the development of classical Petschek
switch-off-slow shocks (SSS). The shock structure that does develop is
controlled by the firehose stability parameter
$\varepsilon=1-\mu_0(P_\|-P_\perp)/ B^2$ through its influence on the 
speed order of the intermediate and slow waves. Here $P_\|$ and $P_\perp$ are
the pressure parallel and perpendicular to the local magnetic
field. The exhaust boundary is made up of a series of two shocks and a
rotational wave. The first shock takes $\varepsilon$ from unity upstream
to a plateau of $0.25$ downstream. The condition $\varepsilon =0.25$ is
special because at this value the speeds of nonlinear slow and
intermediate waves are degenerate. The second slow shock leaves
$\varepsilon=0.25$ unchanged but further reduces the amplitude of the
reconnecting magnetic field. Finally, in the core of the exhaust
$\varepsilon$ drops further and the transition is completed by a rotation
of the reconnecting field into the out-of-plane direction. The
acceleration of the exhaust takes place across the two slow shocks but
not during the final rotation. The result is that the outflow speed
falls below that expected from the Wal\'en condition based on the
asymptotic magnetic field. A simple analytic expression is given for
the critical value of $\varepsilon$ within the exhaust below which SSSs
no longer bound the reconnection outflow.
\end{abstract}

\maketitle

\section{Introduction}

Magnetic reconnection plays an important role in converting magnetic
energy to plasma kinetic and thermal energy. The bulk of the energy
released during reconnection takes place downstream of the magnetic
X-line where reconnected field lines expand outwards to release their
tension. Further, the geometry of the outflow exhaust ultimately
determines the rate of reconnection, the narrow exhaust of the
Sweet-Parker model producing slow reconnection and the open outflow
exhaust proposed by Petschek producing fast reconnection
\cite{petschek64a}. Petschek specifically proposed that a pair of
switch-off slow shocks (SSS) stand in the inflow into the exhaust and
convert the released magnetic energy into the Alfv\'enic outflow
and thermal energy. Magnetohydrodynamic simulations of reconnection
with a localized resistivity imposed at the X-line confirmed the
development of Petschek's open outflow configuration and the formation
of SSSs at the exhaust boundaries \cite{sato79a}. Particle heating by
these shocks has been proposed to cause the X-ray emission in
solar flares \cite{tsuneta96a,longcope11a}.

Magnetic reconnection also takes place in environments in which collisions
are either weak or essentially absent. Observations of reconnection in
the solar wind in particular seem to suggest that reconnection X-lines
and associated exhausts grow to very large scales and resemble the
open outflow geometry predicted by Petschek \cite{phan06a}. On the
other hand, direct observations of SSSs in the Earth's
magnetosphere and the solar wind are infrequent \cite{seon96a}. 

An important result of the simulations of collisionless reconnection
and the resultant Hall reconnection model is the spontaneous formation
of an open outflow configuration analogous to that predicted by
Petschek. This configuration develops even without an ${\it ad hoc}$
localized resistivity \cite{kleva95a,ma96a}. A key question is whether
the fast rates of reconnection seen in the relatively small systems
explored in simulations persist in very large systems. Scaling studies
seem to suggest that collisionless reconnection remains fast in very
large systems \cite{shay99a,shay04a,shay07a}. On the other hand, while
Hall MHD simulations of collisionless reconnection reveal the
formation of SSSs at the boundaries of reconnection exhausts, kinetic
simulations (hybrid or PIC models) reveal that the reconnecting
magnetic field never switches off as expected from the SSS model
\cite{lottermoser98a}. Since it is the release of magnetic energy
downstream from the X-line that ultimately drives the outflow rather
than the dynamics close to the X-line, the absence of the SSS in
kinetic simulations calls into question the conjecture that fast
collisionless reconnection actually can scale to very large
systems. Thus, a key requirement for demonstrating that Hall
reconnection can explain the fast release of energy that takes place
in large systems is to pin down the specific mechanism driving the
Alfv\'enic outflow.

It is well known from observations \cite{hoshino98b,gosling05a,phan07a}
and modeling \cite{hoshino98b,arzner01a,drake09a} of anti-parallel
magnetic reconnection that ions counterstream through the exhaust with
relative velocities of the order of the Alfv\'en speed, increasing the
plasma pressure parallel to the local magnetic field. The consequence
of this increase in parallel pressure is the development of a barrier
in the pseudo-potential that describes the slow shock
transition. This prevents the slow shock from switching-off the
reconnecting magnetic field \cite{yhliu11a,yhliu11b}. For highly
oblique slow shocks relevant to reconnection and where the upstream value of $\varepsilon$ is unity the critical value of the
firehose stability parameter $\varepsilon_{cr}$ below which the SSS
solution becomes inaccessible is given by
\begin{equation}
\varepsilon_{cr}=\frac{5\beta_u+2}{5\beta_u+5},
\label{varepsiloncrit}
\end{equation}
where $\beta_u$ is the ratio of the plasma to magnetic pressure
upstream. If $\varepsilon$ falls below this critical value anywhere
within the exhaust, the SSS solution becomes inaccessible.
$\varepsilon$ does fall below this value in kinetic reconnection
simulations \cite{yhliu11a}. The presence of the potential barrier to
the formation of the SSS actually arises because of the formation of a
new slow shock solution, which is characterized by intermediate Mach
number $M_{I}=\sqrt{V_x^2\mu_0\rho/\varepsilon B_x^2}=1$ both upstream
and downstream \cite{yhliu11b}. We denote this as a 1,1 slow shock or
11-SS. This 11-SS can also switch off the magnetic field downstream
but differs from the SSS where the downstream intermediate Mach number
$M_{Id}<1$. However, we show in the present manuscript that this shock
solution cannot reduce $\varepsilon$ below the critical value given in
Eq.~(\ref{varepsiloncrit}) and therefore can not explain the values of
$\varepsilon$ below $\varepsilon_{cr}$ seen in the reconnection
simulations \cite{yhliu11a}.

Riemann simulations of the structure of slow shocks
\cite{yhliu11a}, in which a constant normal magnetic field is added to
a Harris equilibrium \cite{lin93a,scholer98a}, have revealed that the
value of $\varepsilon$ downstream of the shock transition tends to form a
plateau at a value of $0.25$ and it was demonstrated that at this
value the phase speeds of nonlinear intermediate and slow waves are
degenerate \cite{yhliu11b}. It was suggested that the coupling to a
rotational mode at $\varepsilon=0.25$ produced turbulence at the ion
inertial scale $d_i=c/\omega_{pi}$ that produced sufficient scattering
of ions to prevent $\varepsilon$ from falling below $0.25$. However,
neither the SSS nor the 11-SS can reduce $\varepsilon$ from unity upstream
to $0.25$ downstream.

Here we present the results of a PIC reconnection simulation that is
large enough to reveal the formation of an $\varepsilon=0.25$
plateau. The structure of the shocks that make up reconnection exhausts
are explored by comparing the results of these reconnection
simulations with parallel Riemann simulations and analytic analysis
based on the anisotropic MHD equations. We show that the exhaust
boundary is defined by two slow shock transitions followed by an RD. An
anomalous slow shock (A-SS) \cite{karimabadi95a}, in which $M_{I} \gtrsim 1$
both upstream and downstream, reduces the upstream tangential magnetic
field and takes $\varepsilon$ from unity upstream to $0.25$. A second
slow shock forms on the $\varepsilon=0.25$ plateau and further reduces
the tangential magnetic field. This degenerate slow shock (D-SS) is a special case of the 11-SS 
corresponding to $M_I=M_{SL}=1$ both upstream and downstream, where $M_{SL}$
is the slow mode Mach number. The D-SS forms a compound wave with a
RD whose rotation onsets when $\varepsilon$ begins to fall below
$0.25$. In the reconnection simulation the core of the exhaust becomes
firehose unstable, which prevents further acceleration of the exhaust
outflow by the RD. As a consequence, the exhaust velocity falls below
the value expected based on the upstream Alfv\'en speed.

\section{Simulation details and results}
The PIC simulation is performed with the $P3D$ code \cite{zeiler02a} with
periodic boundary conditions. The initial state is a double-Harris
configuration. However we focus on one current sheet of
$B_z=-B_a\mbox{tanh}(x/w_i)$ and $n=n_h\mbox{sech}^2(x/w_i)+n_a$.  The
magnetic field is normalized to the asymptotic magnetic field $B_a$,
the density to the Harris density $n_h$, velocities to the Alfv\'en
speed $C_A\equiv B_a/\sqrt{\mu_0 m_i n_h}$, lengths to the ion
inertial length $d_i \equiv \sqrt{m_i/\mu_0 n_h e^2}$, times to the
inverse ion cyclotron frequency $\Omega_{ci}^{-1}\equiv m_i/B_ae$, and
temperatures to $m_i C_{A}^2$.  Other important parameters are
$m_i/m_e= 25$, $c=15$, $n_a=0.2$, $w_i=2$ and 
uniform initial $T_{i,e}=0.25$, which imply that
$\beta_a=0.2$. The system size is $819.2 d_i\times 409.2 d_i$ resolved
by grids $16384\times8192$ with $100$ particles per cell. Particles
are advanced with a time step $\Delta t=0.01$.

\begin{figure}
\includegraphics[width=9cm]{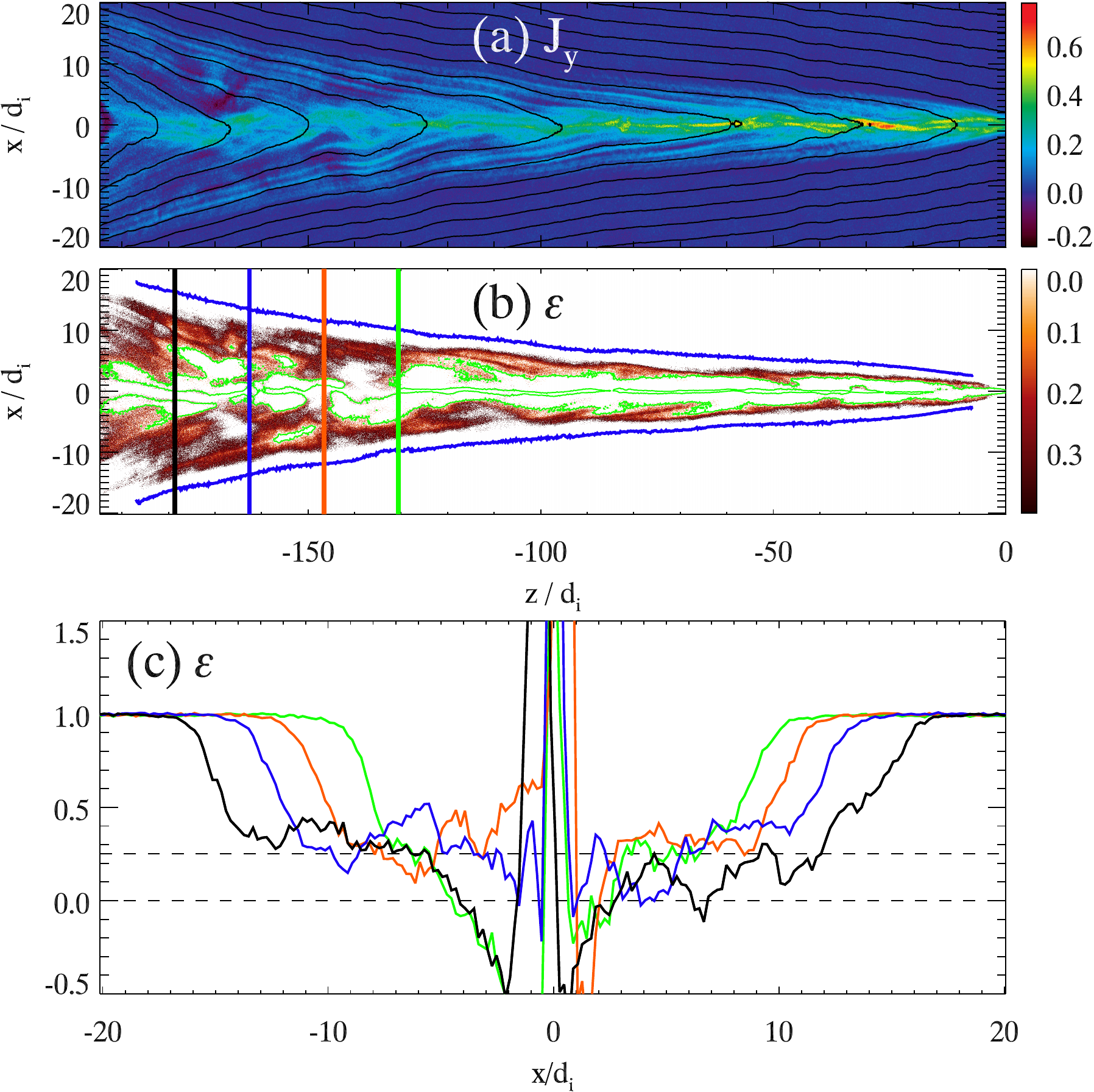} 
\caption{ In (a) the out-of-plane current density $J_y$ overlaid with
  the in-plane magnetic field. In (b) the pair of blue curves bound the
  region $\varepsilon<1$, the red colored areas have
  $0.1<\varepsilon<0.4$ and the green curves bound regions
  with $\varepsilon< 0.0$. In (c) are cuts of $\varepsilon$ at four
  locations marked in (b). Note the tendency to form a plateau at
  0.25. }
\label{LO_1}\end{figure}

In Fig.~\ref{LO_1}(a) is the out-of-plane current density $J_y$ from a
simulation with quasi-steady reconnection. The X-line is at the
origin. The black contours are the in-plane magnetic field lines. The
outflow exhaust forms a Petschek-like open-outflow configuration. In
(b) we show the firehose stability parameter $\varepsilon$. The
counterstreaming ions that have been widely documented in satellite
observations of reconnection exhausts
\cite{hoshino98b,gosling05a,phan07a} drive the anisotropy toward the
firehose threshold. The blue curves bound the region where
$\varepsilon < 1$. The red color table marks the region where
$\varepsilon$ falls between $0.1$ and $0.4$. In the central region of
the exhaust, marked by the green contour, is the firehose unstable
region where $\varepsilon < 0.0$. The significant structure in the
firehose unstable region suggests that this instability is active in
the core of the exhaust \cite{lottermoser98a, karimabadi99a}. With increasing distance from the X-line the
region where $0.1<\varepsilon <0.4$ widens. In (c) we show cuts of
$\varepsilon$ across the exhaust at four locations as marked in
(b). The tendency to form a plateau at $0.25$ is evident in spite of
the turbulence from the firehose unstable region.  Petschek's
reconnection model requires the development of a standing SSS that
propagates in the inflow frame at the phase speed of intermediate
wave.  Since the intermediate wave speed goes to zero when $\varepsilon$
goes to zero, the formation of a slow shock at the exhaust boundary is
going to differ substantially from the traditional SSS. Similar
plateaus at $\varepsilon =0.25$ have been identified in Riemann
simulations of slow shocks \cite{yhliu11a}.

The structure of the exhaust and its boundaries are shown in greater
detail in Fig.~\ref{LO_2} where various physical quantities are
plotted along the black cut in Fig.~\ref{LO_1}(b). The gradual drop in
$B_z$ (in (b)) beginning at the exhaust boundary and the corresponding
rise in pressure (in (d)), with the total plasma and magnetic pressure
(in (d)) being nearly constant, is clear evidence for the development
of a slow shock at the exhaust boundary. However, the downstream $B_z$
does not appear to be switched-off (${\it i.e.}$, $B_z$ does not drop
to zero and remain zero in the core of the exhaust) as in a MHD
SSS. The out-of-plane component $B_y$ indicates the presence of the
residual of the right-handed polarized dispersive whistler wave that
coincides with the front of the SS transition. These whistler waves open
the reconnection dissipation region into an Petschek-like
configuration \cite{drake08a}. Near the center of the exhaust, $B_y$
increases in a left-handed (LH) polarization sense, which is the
expected signature of an intermediate mode. However a clear LH
polarized intermediate mode has not developed.  The outflow $V_z$ (in
(g)) is driven by the ${\bf J}\times {\bf B}$ force (magnetic tension)
associated with the decrease in the magnetic field $B_z$ and is linked
to the jump in $B_z$ by the Wal\'en relation, ${\bf V}_{zd}-{\bf
  V}_{zu} = \pm \sqrt{\rho_u \varepsilon_u/\mu_0}({\bf
  B}_{zd}/\rho-{\bf B}_{zu}/\rho_u)$, where the subscripts ``u'' and
``d'' indicate upstream and dowstream values of the parameters,
respectively \cite{sonnerup81a}. The Wal\'en prediction matches the
exhaust velocity very well (dot-dashed blue curve in (g)) in the regions
where $B_z$ decreases gradually outside of the core of the
exhaust. However, contrary to the Wal\'en prediction, no further
increase in the outflow velocity occurs when $|B_z|$ decreases sharply
to zero in the center of the exhaust. This is probably because the
magnetic tension in this firehose unstable region is zero. The
consequence is that the reconnection outflow speed is typically $\sim
40\%$ slower than the Alfv\'en speed based on the asymptotic magnetic
field.

In order to get a better idea of how slow shocks and rotational waves
propagate out from the central exhaust to a much larger distance
($\sim 100 d_i$) than can be achieved in a reconnection simulation, we
compare the results of Fig.~\ref{LO_2} with those from a quasi-1D
Riemann problem in Fig.~\ref{LO_2_75d}. The angle $\theta_{BN}$
between the upstream magnetic field and the shock normal for this
simulation was taken to $75^\circ$, which is the approximate value
just upstream of the exhaust boundary in Fig.~\ref{LO_2}(f). Data from
this simulation was presented earlier \cite{yhliu11a}. The cuts are
very similar to those obtained from the cuts across the reconnection
exhaust but there are differences. In Fig.~\ref{LO_2_75d}(a) the
plateau in $\varepsilon$ at $0.25$ is much more developed and the
region of firehose instability is much smaller than in the
reconnection simulation.
Unlike the results from reconnection, the downstream
LH rotational wave in Fig.~\ref{LO_2_75d}(b) has clearly
developed. These rotational waves have been identified as the downstream
dispersive wavetrains of slow shocks \cite{coroniti71a}. However, the
wavelength of the oscillations has been shown to depend on the shock
transition thickness, which is not consistent with the dispersive
wavetrain hypothesis \cite{yhliu11a}. Finally in the Riemann solution
the outflow continues to be linked to the variation in $B_z$,
consistent with the Wal\'en relation (see the dot-dashed blue curve in
Fig.~\ref{LO_2_75d}(g)). Again, this difference is likely because
$\varepsilon$ does not fall below the marginal firehose stability threshold in
the Riemann simulation.

\section{1-D shock solutions}
To understand the structure of the shocks that appear in the
reconnection and Riemann simulations, we evaluate the possible
transitions using the anisotropic Rankine-Hugoniot jump
conditions. These follow from the moment integration of the Vlasov
equation for a monatomic plasma assuming that the off-diagonal
pressure components are negligible \cite{chao70a, hudson70a,
  yhliu11b}:
\begin{equation}
\left[ \rho V_x \right]^u_d=0
\label{continue_J}
\end{equation} 
\begin{equation}
\left[\rho
V_x^2+P+\frac{1}{3}\left(\varepsilon+\frac{1}{2}\right)\frac{B^2}{\mu_0}-\varepsilon
\frac{B_x^2}{\mu_0} \right]^u_d=0
\label{normal_m_J}
\end{equation} 
\begin{equation}
\left[\rho V_x {\bf V}_t -\varepsilon \frac{B_x{\bf B}_t}{\mu_0} \right]^u_d=0
\label{transverse_m_J}
\end{equation} 
\begin{equation}
\begin{split}
\left[\left(\frac{1}{2} \rho V^2 + \frac{5}{2}P
+\frac{1}{3}(\varepsilon-1)\frac{B^2}{\mu_0}\right)V_x-(\varepsilon-1)
\frac{B_x{\bf B}_t}{\mu_0} \cdot {\bf V}_t-(\varepsilon-1)
\frac{B_x^2}{\mu_0}V_x +Q_x \right]^u_d=0
\label{energy_J}
\end{split}
\end{equation}

$[...]^u_d$ indicates the difference between up- and downstream. The
equations are written in the deHoffmann-Teller frame where the electric
field vanishes, ${\bf E}=V_x {\bf B}_t- {\bf V}_t B_x=0$. Quantities
$\rho$, $V_x$, ${\bf V}_t$, $B_x$, ${\bf B}_t$ and $Q_x$ are the mass
density, velocity of the bulk flow in the normal direction
($\hat{x}$), velocity of the bulk flow in the tangential direction
($y$-$z$ plane), normal component of the magnetic field, tangential
components of the magnetic field, and the heat flux in the
$x$-direction. Here $P \equiv (P_{||}+2P_\perp)/3$. Even with a known
upstream state, these jump relations still have free parameters
$\varepsilon_d$ and $Q_{xd}$ for which we will use the measured values
from the simulations. The observed $Q_x$ does not jump across the SS
transition of interest so we can discard it.

The anisotropic MHD equations can be further simplified in the case
relevant for magnetic reconnection where the normal magnetic field
$B_x$ is much smaller than the upstream transverse field ${\bf
  B}_{tu}$. In this case the normal velocity $V_x$ is of the order of
$C_{Ax}=B_x/\sqrt{\mu_0\rho}$ and is therefore also small compared with
the fast mode wave speed. The fast mode can then be eliminated from the equations and to lowest order. Eqs.~(\ref{continue_J})-(\ref{energy_J}) become
\begin{equation}
\left[\Gamma\right]^u_d=0,
\label{continue_red}
\end{equation} 

\begin{equation}
\left[P+\frac{1}{3}\left(\varepsilon+\frac{1}{2}\right)\frac{B_t^2}{\mu_0} \right]^u_d=0,
\label{normal_m_red}
\end{equation}

\begin{equation}
\left[\Gamma^2\frac{{\bf B}_t}{\rho} -\varepsilon \frac{B_x^2{\bf B}_t}{\mu_0} \right]^u_d=0,
\label{transverse_m_red}
\end{equation}
 
\begin{equation}
\left[\frac{1}{2} \Gamma^2\frac{B_t^2}{\rho^2B_x^2} + \frac{5}{2}\frac{P}{\rho}-\frac{2}{3}(\varepsilon-1)\frac{B_t^2}{\mu_0\rho} \right]^u_d=0,
\label{energy_red}
\end{equation} 
where the constant mass flux has been written as $\Gamma=\rho V_x$ and
${\bf V}_t$ has been eliminated from the zero constraint on the
electric field. 

We first address why a SSS solution does not appear in either the
reconnection or Riemann solutions. Since there is no reliable analytic
model for the downstream value of $\varepsilon$, we use the values of
$\varepsilon$ from the simulations to explore the possible shock
transitions. In Fig.~\ref{fig4}(a), we plot $\varepsilon$ versus $B_z$
from the right half part of the $75^\circ$ Riemann simulation and a
similar simulation with $\theta_{BN}=30^\circ$ (black curves). In the
$30^\circ$ data $B_z$ decreases steadily from its upstream value to
zero. The SS transition in this simulation corresponds to a SSS in
which $B_{zd}=0$ \cite{yhliu11a}. In the $75^\circ$ data $B_z$
decreases down to around $0.67$ where $\varepsilon=0.25$ and then
decreases further on the plateau where $\varepsilon$ remains at
$0.25$. In this case, as we have discussed earlier, there is no SSS
transition. To understand the reason for these differences, we study
the SS solutions from Eqs.~(\ref{continue_red})-(\ref{energy_red}). For
$\varepsilon_u=1$ and $B_{zu}=1$ we find the traditional SSS solution in
which the upstream intermediate Mach number $M_{Iu}\equiv
V_{xu}\sqrt{\mu_0\rho_u}/B_x$ is unity and the density compression
across the shock $R_\rho\equiv\rho/\rho_u  $, downstream transverse flow $V_{zd}$, and Mach number $M_{Id}$ are given by
\begin{equation}
R_\rho = \frac{\beta_u}{2/5+\beta_u},
\label{SSSRrho}
\end{equation}
\begin{equation}
V_{zd}=\frac{B_{zu}}{\sqrt{\mu_0\rho_u}},
\end{equation}
\begin{equation}
M_{Id}^2=\frac{1}{\varepsilon_dR_\rho},
\end{equation}
where $\beta_u$ is the upstream value of $\beta=2P_u/\mu_0B_z^2$. Note
that the solution is independent of the downstream value of
$\varepsilon$. $M_{Id}$ depends on $\varepsilon_d$ only because the
intermediate wave speed depends on this parameter. 

The disappearance of the SSS solutions is due to the emergence of a
new SS solution from Eqs.~(\ref{continue_red})-(\ref{energy_red}),
an $M_{Iu}=M_{Id}=1$ transition that prevents the SSS solution from
reaching $B_{zd}=0$ \cite{yhliu11b}. Eqs.~(\ref{continue_red})-(\ref{energy_red}) yield the jump conditions on the density and magnetic field,
\begin{equation}
R_\rho=\frac{1}{\varepsilon_d},
\label{compression}
\end{equation}
\begin{equation}
R_B=\frac{B_{zd}}{B_{zu}}=\frac{\frac{5}{2}\varepsilon_d-1-\frac{5}{2}(1-\varepsilon_d)\beta_u}{\frac{1}{3}\varepsilon_d(1+\frac{7}{2}\varepsilon_d)}.
\label{11-SSRB}
\end{equation}
 This 11-SS solution is shown in the red lines in Fig.~\ref{fig4}(a)
 as obtained from Eqs.~(\ref{continue_J})-(\ref{energy_J}). Note that
 the 11-SS does not intersect the measured $B_z-\varepsilon$ curve for
 the $30^\circ$ case at a point other than the upstream state, but
 does for $75^\circ$. If the 11-SS solution intersects the
 $B_z-\varepsilon$ curve the SSS solution is no longer
 accessible. This is because the presence of the 11-SS solution at a
 finite value of $B_{zd}$ signifies the formation of a minimum in the
 effective potential $\psi(B_y,B_z)$ that describes the SSS transition
 and a corresponding maximum at the origin where $B_z=B_y=0$
 \cite{yhliu11b}. The potential in this case is similar to that shown
 in Fig.~\ref{fig5}(b). The survival of the SSS solution therefore
 requires that the minimum of $\varepsilon_d$ as determined from
 either simulation or observational data fall above the minimum of
 $\varepsilon$ that is accessible by the 11-SS. This minimum is
 obtained from Eq.~(\ref{11-SSRB}) by taking $B_{zd}=0$. Thus, the SSS
 can form as long as the minimum of $\varepsilon$ across the outflow
 exhaust exceeds
\begin{equation}
\varepsilon_{cr}=\frac{5\beta_u+2}{5\beta_u+5}.
\label{varepsilon-cr}
\end{equation}
Thus, in the case of anti-parallel reconnection where counterstreaming
ions reduce $\varepsilon$ below this critical value there is no SSS
transition. On the other hand, it is also evident from
Fig.~\ref{fig4}(a) that the 11-SS can not produce a transition below
the minimum value given in Eq.~(\ref{varepsilon-cr}), which lies above
the $0.25$ measured in the simulations.

The only option remaining for finding a SS solution that can
transition to $\varepsilon_d=0.25$ is to increase the value of the
upstream Mach number $M_{Iu}$ above unity. That increasing $M_{Iu}$
facilitates a SS transition to a lower value of $\varepsilon_d$ and is
shown in Fig.~\ref{fig4}(b), which shows the $B_z-\varepsilon$
relation obtained from Eqs.~(\ref{continue_J})-(\ref{energy_J}) for
$M_{Iu}=1.29$. The value of $M_{Iu}$ was increased until the shock
transition curve intersected the black curve at the red circle
($B_{z,d}=0.67, \varepsilon_d=0.25$). This super-intermediate SS
transition, which was earlier denoted as an anomalous slow shock
(A-SS) \cite{karimabadi95a} is a possible solution for the SS seen in
the $75^\circ$ Riemann simulation. A similar solution with
$M_{Iu}=1.17$ that results in an intersection point ($B_{z,d}=0.65,
\varepsilon_d=0.25$) is shown with the data from the reconnection
simulation in Fig.~\ref{fig4}(c).

\begin{figure}
\includegraphics[width=9cm]{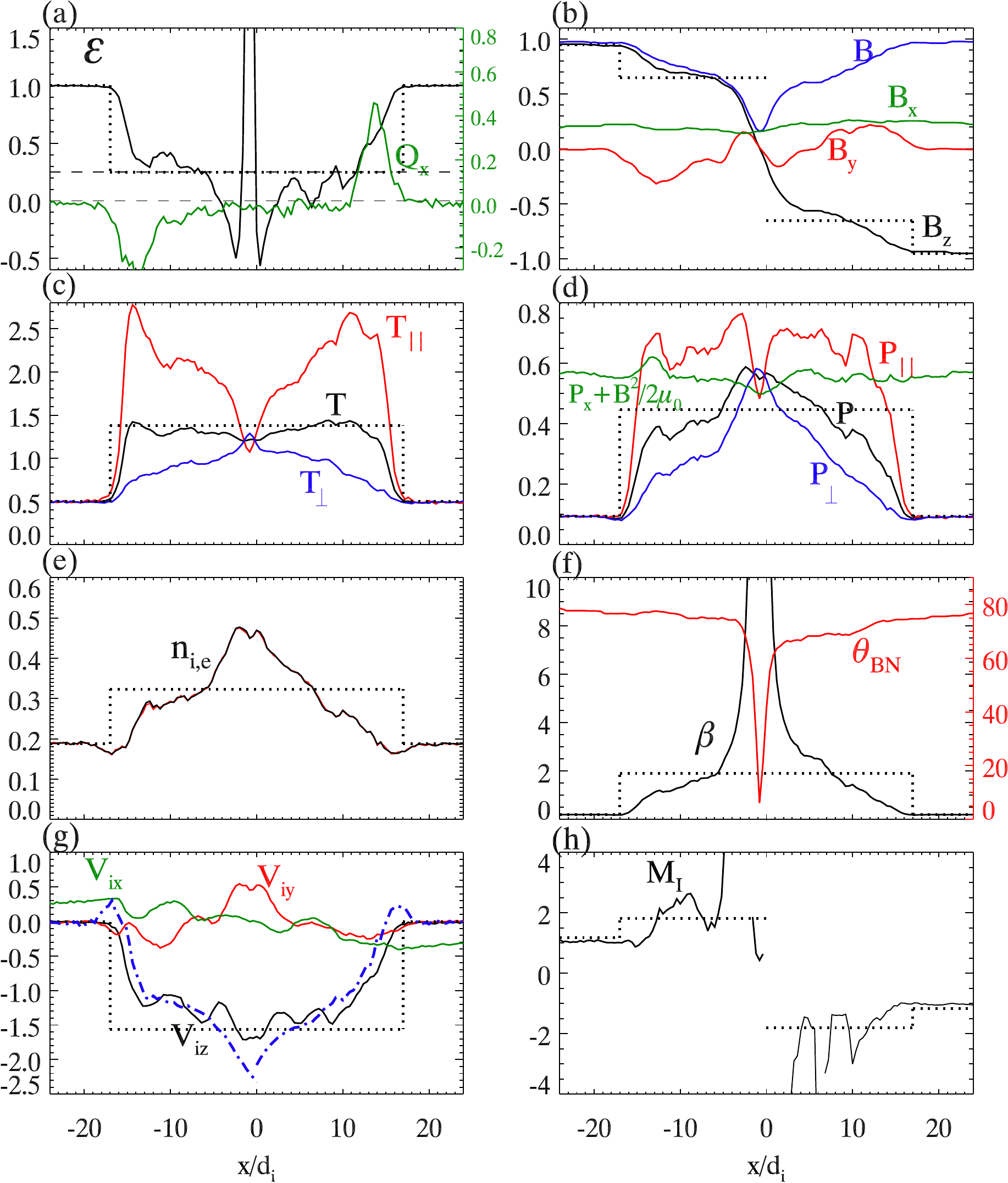} 
\caption{From the reconnection simulation profiles of physical
  quantities along the black cut of Fig.~\ref{LO_1}(b). $\varepsilon$
  and $Q_x$ in (a), $B$'s in (b), $T$'s in (c), $P$'s in (d), $n_i$
  and $n_e$ in (e), $\beta$ and $\theta_{BN}$ in (f), $V$'s in (g) and
  $M_I$ in (h). $\theta_{BN}$ is the angle between the local magnetic
  field and the x-direction. The dot-dashed blue curve in (g) is the
  Wal\'en relation. The dotted black curves are the predicted jumps of
  an anomalous slow shock with $M_{Iu}=1.17$, and these dotted curves 
  should be compared with the solid black curves.}
\label{LO_2}\end{figure}

\begin{figure}
\includegraphics[width=9cm]{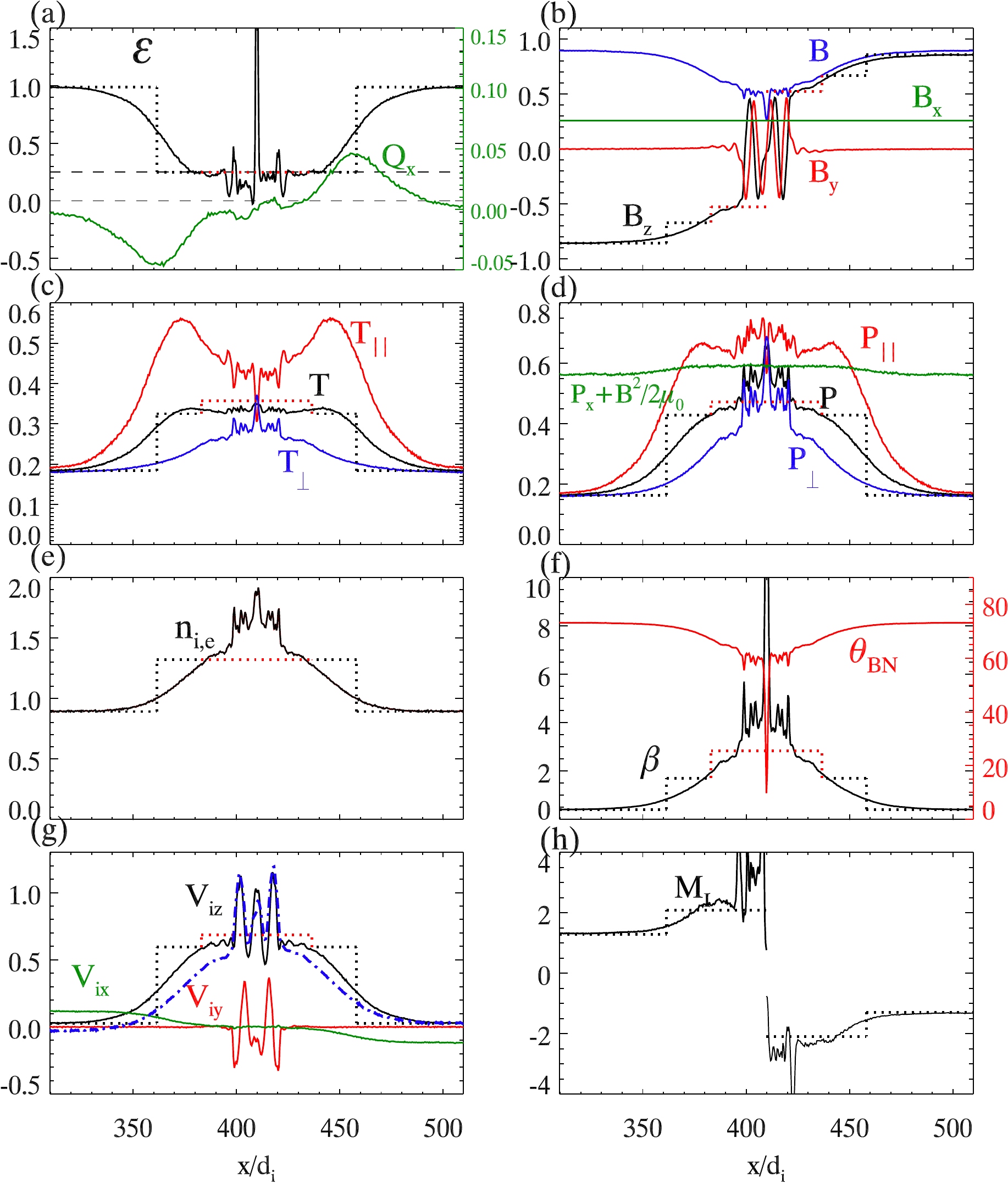} 
\caption{Profiles of physical quantities of Riemann $75^\circ$ case
  with system size $1.6 d_i \times 816.2 d_i$ at time
  $200/\Omega_{ci}$ \cite{yhliu11a} presented in the same format as
  Fig.~\ref{LO_2}. The dotted black and red curves are the predicted jumps of
  an A-SS with $M_{Iu}=1.29$ followed by the D-SS. Both dotted curves should be compared with the solid black curves.}
\label{LO_2_75d}\end{figure}

\begin{figure}
\includegraphics[width=16cm]{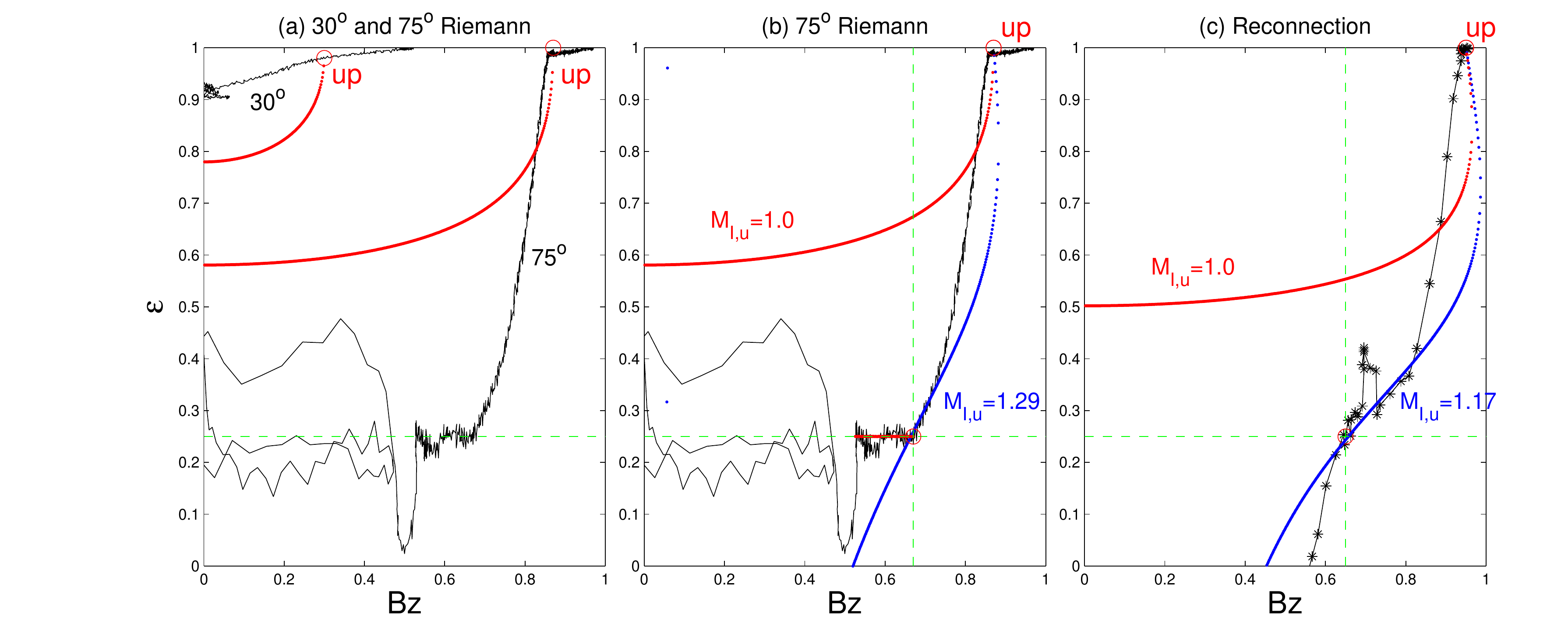} 
\caption{ The $B_z-\varepsilon$ space where the downstream direction
  is toward $B_z=0$. In (a) the lower black curve is measured from the
  right half part of data from the Riemann simulation in Fig.~\ref{LO_2_75d}
  and the upper curve is from a similar simulation with
  $\theta_{BN}=30^\circ$. The red curves show possible transitions of
  a $M_{Iu}=M_{Id}=1$ slow shock transition (11-SS) from the
  anisotropic MHD equations. If the red 11-SS curve intersects the
  black curve at any location other than in the upstream state, the
  switch-off-slow shock transition is not possible. In (b) the
  $75^\circ$ Riemann data in black, the 11-SS transition in red and a
  $M_{Iu}=1.29$ SS transition (A-SS) in blue. The intersection of
  the black and the blue curves at the red circle $(B_z=0.67,
  \varepsilon=0.25)$ is the chosen A-SS downstream since this solution
  allows a transition from $\varepsilon$ unity upstream to a downstream
  value of $0.25$. In (c) plots from the reconnection simulation similar to those in (b) for 
  the Riemann $75^\circ$ case. The red circle $(B_z=0.65,
  \varepsilon=0.25)$ is the chosen downstream state.}
\label{fig4}\end{figure}

\begin{figure}
\includegraphics[width=18cm]{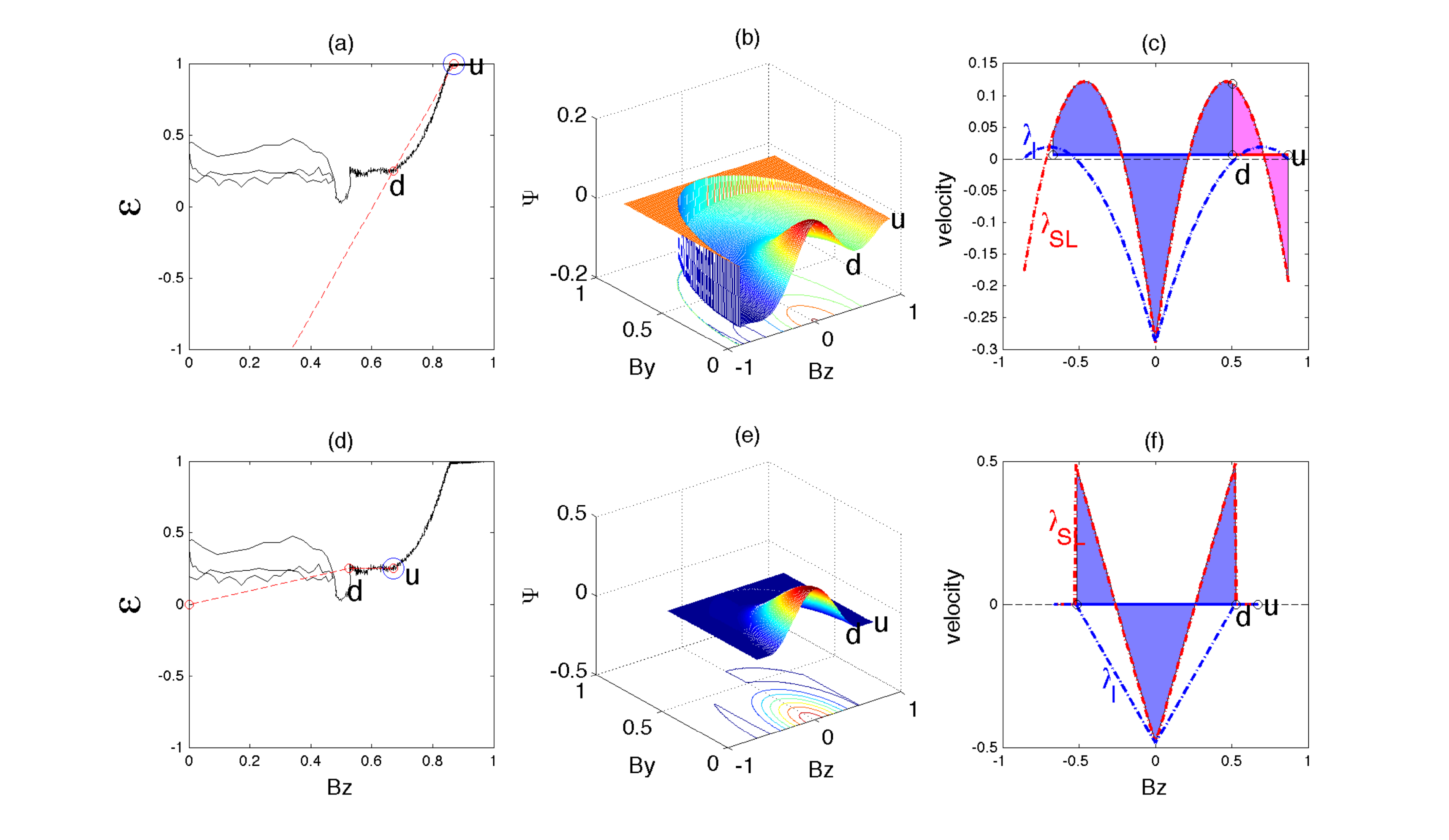} 
\caption{ Analysis of the $75^\circ$ Riemann simulation. In (a) the
  $B_z-\varepsilon$ data from the simulation with a chosen analytic
  relation for $\varepsilon (B_z)$ linking the upstream and downstream
  states. In (b) the pseudo-potential $\Psi$ from anisotropic derivative
  nonlinear-Schr\"odinger-Burgers equations \cite{yhliu11b} showing the
  potential minimum at the downstream solution and the barrier
  preventing the development of the SSS. In (c) the shock speed (red
  solid) and the slow (dot-dashed red) and intermediate (dot-dashed blue) characteristics, 
  $\lambda_{SL}$ and $\lambda_I$ respectively, show that
  the A-SS solution is a super-intermediate to super-intermediate
  transition but a super-slow to sub-slow transition. In (d) similar
  to (a) but with an analytic relation for $\varepsilon (B_z)$
  describing the $\varepsilon=0.25$ plateau region and extension
  toward $\varepsilon=0$. In (e) the pseudo-potential for the D-SS
  showing that the potential is flat between the up and downstream
  states and the barrier that facilitates a left hand rotation that
  describes the compound D-SS/RD transition. In (f) wave
  characteristics for the D-SS/RD as shown in (c). }
\label{fig5}\end{figure}

To further test the hypothesis that the A-SS describes the
reconnection exhaust boundary, we plot the downstream predictions of
the A-SS transition based on Eqs.~(\ref{continue_J})-(\ref{energy_J})
on top of the data from the Riemann and reconnection simulations in
Fig.~\ref{LO_2} and \ref{LO_2_75d}. The predicted jumps of the A-SSs
appear as dotted black lines, and they should be compared with the
solid black lines. Overall, the predicted jumps of these A-SS's agree
very well with the observed transitions.  With the measured inflow
speeds $V_{ix}$, which are $\sim 0.3$ and $\sim 0.11$ in Figs.~\ref{LO_2}(g)
and \ref{LO_2_75d}(g), respectively, we can calculate the shock speeds
of these A-SS's in the rest frame of the far upstream plasma. The local
intermediate Mach number $M_{I}$ is then derived and plotted in
(h). The choice of $M_{Iu}$ being slightly greater than unity and the
shock therefore being super-intermediate is consistent with both the
reconnection and Riemann simulations but is most evident in the
Riemann case. The measured downstream Mach number in Fig.~\ref{LO_2}
(h), although oscillating, is comparable to the prediction, while it
agrees very well in Fig.~\ref{LO_2_75d} (h). Our interpretation of why
the shock is super-intermediate is that the ions stream along the
local magnetic field and into the upstream at the Alfv\'en speed while
encountering a weak inflowing plasma. The interaction between these
two oppositely moving constituents would naturally drive the formation
of a weak super-intermediate slow shock rather than the conventional
standing slow shock.

A better understanding of the A-SS transition can be obtained from the
pseudo-potential that describes the shock transition. This potential
describes the variation of ${\bf B}_t$ as a function of the space
variable in the 2-D $B_y-B_z$ plane. In analogy with 2-D particle
dynamics, the space variable can be treated as time and the $B_y-B_z$
plane can be treated as a 2-D configuration space. We calculate the
potential using the anisotropic derivative
nonlinear-Schr\"odinger-Burgers equation \cite{yhliu11b}. To do this
we need an analytic relation between $B_z$ and $\varepsilon$. For
simplicity we use a straight line from the upstream-to downstream
states shown in Fig.~\ref{LO_2_75d}(b) (Fig.~\ref{fig5}(a)).  The
resulting pseudo-potential is shown in Fig.~\ref{fig5}(b). In this
potential a pseudo-particle starts at an upstream state $B_{zu}$ and
slides down the hill to $B_{zd}$ with $B_y=0$.  The peak in the
potential around $B_z=0$ is evident. This potential hill prevents the
A-SS from making a transition to $B_z=0$. The pseudo-particle could,
in principle, follow the valley by rotating $B_z$ into $B_y$. However,
as discussed below, this rotation does not take place at this location
but further downstream.

The A-SS shock speed (solid red) and the variation of the intermediate
$\lambda_I$ (dot-dashed blue) and slow $\lambda_{SL}$ (dot-dashed red)
wave speeds across the transition are shown in Fig.~\ref{fig5}(c). All
velocities are shown in the upstream intermediate frame. Note that
upstream the A-SS is super-intermediate and super-slow (the shock
speed at $B_{zu}$ exceeds the intermediate and slow characteristic
speeds) and that downstream the A-SS is super-intermediate and
sub-slow (the shock speed at $B_{zd}$ exceeds the intermediate but is
below the slow-characteristic speed). Thus, this A-SS is an $M_{Iu}>1$
to $M_{Id}>1$ transition with $M_{Id}>M_{Iu}$ as is evident in panel
(h) of Figs.~\ref{LO_2} and ~\ref{LO_2_75d}.

Two other distinct features of the exhaust boundaries are also evident
in Figs.~\ref{LO_2} and \ref{LO_2_75d}. In the $\varepsilon=0.25$
pedestal region the transverse magnetic field $B_z$ continues to
decrease and the exhaust velocity continues to increase, the latter
being consistent with the Wal\'en prediction. Finally, further towards
the symmetry line the transverse magnetic field $B_z$ abruptly begins
to rotate into the y-direction. The pedestal region with constant
$\varepsilon=0.25$ can again be explored with
Eqs.~(\ref{continue_red})-(\ref{energy_red}). This case is a special case of the 11-SS solution discussed earlier. As in that solution, $M_I=\sqrt{\Gamma^2\mu_0/\varepsilon\rho B_x^2}$ is a constant so that $\rho\propto\varepsilon^{-1}$ is constant. The energy equation then takes the form
\begin{equation}
\varepsilon (\varepsilon-\frac{1}{4})\left[\frac{B_z^2}{\mu_0} \right]^u_d=0.
\label{energy_D-SS}
\end{equation}
This equation reveals the unusual conditions that arise when
$\varepsilon=0.25$. This SS solution can support arbitrary changes in
the transverse magnetic field and associated pressure (determined by
the constancy of the magnetic plus plasma pressure). The existence of
this transition is marked by the short horizontal red segment at 0.25
in Fig.~\ref{fig4}(b). The transition is less evident in Fig.~\ref{fig4}(c)
because of the limited spatial extension in the normal direction of the 
reconnection simulation.
Since $\rho$ is a constant the transverse flow is given by the simplified Wal\'en relation
$[V_z]^u_d=\sqrt{\varepsilon/\mu_0\rho}[B_z]^u_d$.  
The condition of
constant $\varepsilon=0.25$ corresponds to the degeneracy of slow and
intermediate waves \cite{yhliu11b}. We therefore refer to this
solution as a degenerate slow shock (D-SS). Note that an exactly
horizontal transition at $B_z-\varepsilon$ plane ({\it i.e.,} caused
by scattering \cite{yhliu11a,yhliu11b}) can only take place at
$\varepsilon=0.25$ in the form of the D-SS.  The pseudo-potential can
again be constructed for the D-SS. The chosen $B_z-\varepsilon$
relation is shown by the dashed red curve in Fig.~\ref{fig5}(d). It is
a horizontal line from $B_{zu}$ to $B_{zd}$ and then is taken to
$\varepsilon=0$ at $B_z=0$. The latter segment is meant to be
illustrative. The structure of the pseudo-potential and characteristic
speeds shown in Figs.~\ref{fig5}(e-f) are insensitive to the detailed
variation of $\varepsilon$ as long as it drops below
$0.25$. Fig.~\ref{fig5}(e) reveals that the potential is flat on the
$\varepsilon=0.25$ plateau and again has a barrier around the origin
that prevents a direction transition to $B_z=0$. On the plateau the
shock, intermediate and slow wave speeds are equal to the upstream
intermediate speed (Fig.~\ref{fig5}(f)). The predicted weak jumps of
D-SS (dotted red) of the Riemann simulation in Fig.~\ref{LO_2_75d} are
added to jumps of A-SS (dotted black) for further comparison with
data. The agreement between the simulation data and the D-SS
predictions are reasonably good.

The data from both the reconnection and Riemann simulations reveal the
remnant upstream tangential magnetic $B_z$ abruptly begins a LH
rotation at the lower edge of the $\varepsilon$ plateau. This is
especially evident in the Riemann simulations. The transition of the
coplanar D-SS to this LH rotational mode follows from the structure of
the pseudo-potential in Fig.~\ref{fig5}(e). The pseudo-particle moves
toward lower $B_z$ but must rotate as a result of the barrier around
the origin. Fig.~\ref{fig5}(f) reveals that, as expected, the downstream
intermediate characteristic speed decreases with decreasing $\varepsilon$ and it becomes slower than the slow characteristic speed. 
The rotational mode is not an intermediate shock (IS) which must be
super-intermediate upstream and sub-intermediate downstream ({\it i.e.,} the blue portion in Fig.~\ref{fig5}(c)). Thus, the
rotational wave should be interpreted as a rotational discontinuity
(RD) that is linked to the D-SS, forming a compound D-SS/RD. The rate
of rotation in the Riemann simulation is significantly higher than
that in the reconnection simulation. This is likely because
$\varepsilon$ falls to zero in the core of the exhaust in the
reconnection simulation, which reduces the rotation rate of the LH
wave. 

The dynamical picture leading to these multiple structures can be
understood as following: The counter-streaming ions drives the overall
tendency to form a A-SS + RD wave (Fig.~\ref{fig5}(c)). However, due
to rotation-induced scattering the $\varepsilon$ tends to form a
plateau at $0.25$ \cite{yhliu11a,yhliu11b} and the downstream region
locally evolves into a D-SS/RD compound wave (Fig.~\ref{fig5}(f)).

\section{Implication}

For a density of $1.0 cm^{-3}$, typical of the magnetotail, $1d_i=150km$,
which implies that the simulation shown in Fig. 1 corresponds to
roughly $20 R_E$.  During times when the tail has a single large
X-line, the signatures discussed in this paper should be observable by
Geotail and THEMIS. Specifically, we predict that no switch-off-slow
shocks should develop if the measured values of $\varepsilon$ across
the exhaust fall below the threshold $\varepsilon_{cr}$ given in
Eq.~(\ref{varepsiloncrit}). We predict that the exhaust boundary will
exhibit two distinct SS transitions. The first is an anomalous slow
shock (A-SS) that is a super-intermediate to super-intermediate
transition that drops $\varepsilon$ from unity to around $0.25$. The
second is a degenerate slow-shock (D-SS)/rotational-discontinuity (RD) compound wave that
forms on and maintains a plateau with $\varepsilon=0.25$. When
$\varepsilon=0.25$ the slow and intermediate wave speeds are equal and
the resulting D-SS can support an isobaric
change in magnetic and particle pressure while leaving the density
constant. The RD rotates the residual transverse magnetic field in
the left-hand sense. 
Since the cross-exhaust width of the firehose unstable region
does not increase with distance downstream of the X-line in reconnection simulation, we suggest
that the dissipation region proper is the source of the firehose
unstable region. This firehose unstable region causes the failure of Wal\'en relation.


Several caveats must be kept in mind.  First, because of the
relatively narrow angle of the reconnection outflow exhaust, small
reconnection events might be too narrow for the $\varepsilon=0.25$
plateau to develop.  Second, we have only considered 2D simulations
that are invariant in the y direction.  A more realistic 3D system may
develop other features that mask or even change our
conclusions. Finally, unlike the anti-parallel case, with a
significant ambient initial guide field $B_y$ the firehose parameter
$\varepsilon$ does not become as small and as will be shown in a
future manuscript a pair of RD's switch off the reconnecting magnetic
field ($B_z$).

\acknowledgments

Y.\ -H.\ L.\ acknowledges helpful discussions with \ W. \ Daughton,
\ H.\ Karimabadi and \ L.\ Hui. This work was supported in part by
NASA grant Nos. NNX08AV87G, NNX09A102G, NASA's Heliophysics Theory Program and a grant from the DOE/NSF partnership in basic plasma physics. Computations
were carried out at the National Energy Research Scientific Computing
Center. LA-UR-11-12116. \newpage

\appendix{Appendix: Exact solutions of the anisotropic MHD equations}

Equations~(\ref{continue_J})-(\ref{energy_J}) have analytic solutions,
\begin{equation}
A_x^2 \equiv \frac{V_x^2}{B_x^2/(\mu_0 \rho)}=\frac{-b \pm \sqrt{b^2-4ac} }{2a},
\label{Ax2}
\end{equation} 
where
\begin{equation}
a=\frac{2}{5}\frac{B^2}{B_x^2},
\end{equation} 
\begin{equation}
b=-A_{xu}^2-\left[\frac{\beta_u}{2}+\frac{1}{3}(\varepsilon_u+\frac{1}{2})\right]\mbox{sec}^2\theta_u+\varepsilon_u-\left[\frac{3}{5}\frac{B_t}{B_x}-\frac{2}{5}\frac{B_x}{B_t}\right]\mbox{tan}\theta_u(A_{xu}^2-\varepsilon_u)-\frac{1}{10}\frac{B^2}{B_x^2},
\end{equation} 
\begin{equation}
c=\frac{1}{5}\mbox{sec}^2\theta_uA_{xu}^4+\left[\frac{\beta_u}{2}-\frac{4}{15}(\varepsilon_u-1)\right]\mbox{sec}^2\theta_uA_{xu}^2,
\end{equation} 

where $Q_x$ has been neglected. Using the identity
\begin{equation}
\varepsilon=A_x^2-\mbox{tan}\theta_u\frac{B_x}{B_t}(A_{xu}^2-\varepsilon_u)
\end{equation}

with given $\theta_u$, $\beta_u$, $\varepsilon_u$ and $A_{xu}$, we can plot $\varepsilon$ versus $B_t$ and obtain the following jump conditions.

\begin{equation}
M_I=\frac{A_x}{\sqrt{\varepsilon}},
\end{equation} 
\begin{equation}
\frac{\rho}{\rho_u}=\frac{A_{xu}^2}{A_{x}^2},
\label{rho_r}
\end{equation} 
\begin{equation}
\frac{\beta}{\beta_u}=\frac{2B_x^2}{\beta_uB^2}\left\{A_{xu}^2+\left[\frac{\beta_u}{2}+\frac{1}{3}(\varepsilon_u+\frac{1}{2})\right]\mbox{sec}^2\theta_u-\varepsilon_u-A_x^2+\varepsilon\right\}-\frac{2}{3\beta_u}(\varepsilon+\frac{1}{2}), 
\end{equation} 
\begin{equation}
\frac{P}{P_u}=\left(\frac{\beta}{\beta_u}\right)\frac{B^2}{B_x^2}\mbox{cos}^2\theta_u,
\end{equation}
and the downstream transverse velocity measured at upstream plasma frame, 
\begin{equation}
\frac{V_t}{B_x/\sqrt{\mu_0 \rho_u}}=A_{xu}\mbox{tan}\theta_u-\frac{B_t}{B_x}\frac{A_x^2}{A_{xu}}.
\end{equation} 
\newpage

In the very oblique limit, we look for solution of $A_{xu}^2=\varepsilon_u$ and $A_x^2=\varepsilon$ (the 11-SS solution). $B_t=0$ is further imposed, since this point gives us the minimum $\varepsilon_{cr}$ of 11-SS transitions. We get
\begin{equation}
a\sim\frac{2}{5},
\end{equation} 
\begin{equation}
b\sim-\left[\frac{\beta_u}{2}+\frac{1}{3}(\varepsilon_u+\frac{1}{2})\right]\mbox{sec}^2\theta_u,
\end{equation} 
\begin{equation}
c\sim\left\{\frac{1}{5}\varepsilon_u^2+\left[\frac{\beta_u}{2}-\frac{4}{15}(\varepsilon_u-1)\right]\varepsilon_u\right\}\mbox{sec}^2\theta_u.
\end{equation} 
Plug these into Eq.~(\ref{Ax2}), choose the plus sign and expand it, the minimum $\varepsilon_{cr}$ value that a SSS (with a given $\varepsilon_u$ and $\beta_u$) can transition to is obtained
\begin{equation}
\varepsilon_{cr}=A_x^2\sim\frac{c}{|b|}=\frac{-2\varepsilon_u^2+(15\beta_u+8)\varepsilon_u}{10\varepsilon_u+15\beta_u+5}.
\end{equation}
Eq.~(\ref{varepsiloncrit}) is re-derived when $\varepsilon_u=1$,
\begin{equation}
\varepsilon_{cr}=\frac{5\beta_u+2}{5\beta_u+5}.
\end{equation}
We noticed that $\varepsilon_u-\varepsilon_{cr} > 0 $ vanishes when $\varepsilon_u$ approaches value 0.25 from unity. Therefore $\varepsilon=0.25$ is the solution of $\varepsilon_u=\varepsilon_{cr}$. From this we can conclude that a SSS with a $\varepsilon_u > 0.25$ and an arbitrary $\beta_u$ can not transition to $\varepsilon < 0.25$ region. In other words, $0.25$ is an "absolute" barrier across which a SSS with $\varepsilon_u > 0.25$ can not cross.

\newpage





\end{document}